\documentclass[12pt]{iopart}

\usepackage{iopams}

\usepackage{xcolor}
\usepackage{hyperref,graphicx}

\newcommand{\politecnico}{Dipartimento di Fisica - Politecnico di Milano, p.za Leonardo da Vinci 32, 20133 Milano, Italy}
\newcommand{\ifncnr}{Istituto di Fotonica e Nanotecnologie - Consiglio Nazionale delle Ricerche, p.za Leonardo da Vinci 32, 20133 Milano, Italy}

\begin{document}

\title{Projecting light beams with 3D waveguide arrays}

\author{Andrea Crespi$^{1,2}$, Francesca Bragheri$^2$}
\address{$^1$ \politecnico}
\address{$^2$ \ifncnr}
\ead{andrea.crespi@polimi.it}

\begin{abstract}
Free-space light beams with complex intensity patterns, or non-trivial phase structure, are demanded in diverse fields, ranging from classical and quantum optical communications,  to manipulation and imaging of microparticles and cells.
Static or dynamic spatial light modulators, acting on phase or intensity of an incoming light wave, are the conventional choices to produce beams with such non-trivial characteristics.
However, interfacing these devices with optical fibers or integrated optical circuits often requires difficult alignment or cumbersome optical setups. Here we explore theoretically and with numerical simulations the potentialities of directly using the output of engineered three-dimensional waveguide arrays, illuminated with linearly polarized light, to project light beams with peculiar structures. We investigate through a collection of illustrative configurations the far field distribution, showing the possibility to achieve orbital angular momentum, or to produce elaborate intensity or phase patterns with several singularity points. We also simulate the propagation of the projected beam, showing the possibility to concentrate light. We note that these devices should be at reach of current technology, thus perspectives are open for the generation of complex free-space optical beams from integrated waveguide circuits.
\end{abstract}


\section{Introduction}

The ability to shape the wavefront of a light beam, producing non-trivial intensity or phase patterns, is of interest for several applications. In particular, the possibility to exploit the spatial degrees of freedom of a multi-mode field to encode information has attracted increasing research in the recent years, with special interest in orbital angular momentum, both of classical and quantum light states,  for free-space \cite{wang2012,vallone2014, milione2015} or fiber-based communications \cite{bozinovic2013, vanuden2014}. Mode beams with angular momentum \cite{rotationOT} as well as structured light \cite{Dholakia2003,HRD2015} are also exploited to trap and manipulate micro and nanoparticles, or even living cells, through optical forces \cite{Padgett2013,cellrotator2008}. In fact, the possibility to draw complex interference patterns in the propagation provides enhanced control on the particle position, rotation or movement, capabilities of former importance when dealing with the analysis and sorting of single cells.

Bulk optics components, such as phase or intensity masks \cite{wang2012, vallone2014, rotationOT}, as well as dynamic spatial light modulators \cite{milione2015, bozinovic2013, HRD2015}, are the almost exclusive choice to shape the wavefront of such free-space propagating beams. Therefore, the interfacing with fiber networks, integrated optical circuits, or fluidic lab-on-a-chips can pose stability or alignment problems, and limit the effective exploitation of the advantages of these beam shaping methods.

On the other hand, three-dimensional waveguide lattices \cite{szameit2006, szameit2007, rechtsman2012, corrielli2013, crespi2015, caruso2016} have emerged in the last decade as powerful devices to investigate engineered multi-mode fields. Due to the high level of control that can be achieved in such structures, they have been employed to observe fundamental effects such as spatial solitons \cite{szameit2007} or edge states in topological insulators \cite{rechtsman2012}, as well as to simulate a variety of quantum physics phenomena\cite{corrielli2013, crespi2015, caruso2016}. 

While waveguide lattices with a simple structure may be induced by external optical fields in photorefractive materials \cite{fleischer2003}, it was the three-dimensional fabrication capability of femtosecond laser waveguide writing \cite{osellame2012} that fostered much of the work in this field. In fact, this micromachining technique is based on the non-linear absorption of femtosecond pulses, focused in the bulk of a dielectric material: judiciously chosen irradiation parameters result in a permanent refractive index increase, localized in the focus region. Translation of the substrate with respect to the laser beam allows to inscribe with full three-dimensional freedom the desired waveguiding paths. However, up to now, the research was mainly focused on the propagation properties of the light within the waveguide arrays and the far-field emission properties were not specifically considered.

Here, we explore the possibility to use properly engineered waveguide lattices to project free-space light beams with controlled wavefront. We numerically and theoretically investigate, also by using relevant examples, the properties of the emitted light from a pattern of waveguide modes, distributed on a two-dimensional cross-section, as it can be achieved with three-dimensional waveguide arrays. In particular we study the possibility to produce peculiar far-field distributions, and one or multiple phase singularity points. We will also show how to obtain a certain degree of control on the propagation properties of the emitted beam, including focusing.

\section{Models and approximations}
\label{sec:general}

This work investigates the properties of the output emission from waveguide arrays, in which identical single-mode optical waveguides are placed in arbitrary position on a two-dimensional cross-section.  We will assume that it is possible to produce arrays maintaining full control on the intensity of the different modes, as well on their relative phase. While we will not go into details of the design and engineering of the arrays, we will discuss their effective feasibility in general terms in \sref{sec:conclusion}.

We will consider the arrays as illuminated with linearly polarized light, with a fixed polarization direction. Of course, including also the polarization degree of freedom would open even wider possibilities on the achievable output light states, but this would be out of the scope of this preliminary study. We will further restrict our study to the paraxial condition. 

In this framework, it is common to treat the electromagnetic field as scalar and the calculation of the field distribution during the propagation be achieved in terms of scalar diffraction theory \cite{svelto4th, goodman2005}. Considering our case in more detail, the output of an array of identical single-mode optical waveguides may be conveniently modelled with an array of Gaussian modes. If we drop uniform phase terms, the field of a Gaussian mode at a certain propagation coordinate $z$, and as a function of the transverse position $\bi{r}$, reads:
\begin{equation}
g(\bi{r},z) = \frac{w_0}{w} e^{- \frac{|\bi{r}|^2}{w^2}} \, e^{-\iota k \frac{|\bi{r}|^2}{2 R}} \label{eq:gaussiano}
\end{equation}
where the beam radius $w$ and the wavefront curvature radius $R$ depends on $z$ as follows:
\begin{equation}
\begin{array}{ccc}
w = w_0 \sqrt{1+ \left(\frac{\lambda z}{\pi w_0^2} \right)^2} &\qquad&
R = z \sqrt{1+ \left(\frac{\pi w_0^2}{\lambda z} \right)^2}
\end{array}
\label{eq:paramGaussiano}
\end{equation}
Now, the Gaussian modes in the array are assumed to be all identical with regard to $w_0$ (which is placed at the output facet of the array, where we can put the origin $z$=0), but they may have different intensity and phase terms. If $\bi{R}_n$ are the positions of the waveguides in the transverse plane, $I_n$ their intensity and $\alpha_n$ their phase terms, the scalar field configuration of the output of an array of $N$ waveguides can be written as:
\begin{equation}
u(\bi{r},z) = \sum^N_{n=1} \sqrt{I_n} e^{\iota \alpha_n} g(\bi{r} - \bi{R}_n,z)   = g(\bi{r},z) \ast \sum^N_{n=1} \sqrt{I_n} e^{\iota \alpha_n} \delta(\bi{r} - \bi{R}_n)
\label{eq:campoArrayGeneral}
\end{equation}
The above expression can be used to calculate the field distribution at an arbitrary coordinate $z$ within the paraxial approximation, and all numerical simulations shown in this work are based on that.

\section{Polygonal arrays and angular momentum}
\label{sec:polygonal}

Optical beams possessing angular momentum have attracted increasing interest in the recent years, for diverse applications: on the one hand orbital angular momentum is a powerful degree of freedom for encoding information, both classical and quantum \cite{wang2012, vallone2014}; on the other hand the possibility to exert torque on trapped microparticles opens novel possibilities in optical micromanipulation \cite{hong2015}. In this scenario it can be important to produce such vortex beams directly from optical fibers and waveguides, with reduced or no use of bulk optical components, which require critical alignment. Different approaches have been proposed, from ring resonators \cite{cai2012} to circular gratings \cite{doerr2011, su2012}, to hybrid plasmonic waveguides \cite{liang2014}.
\begin{figure*}[p]
\centering
\vspace{1.5cm}
\includegraphics[width=0.8\textwidth]{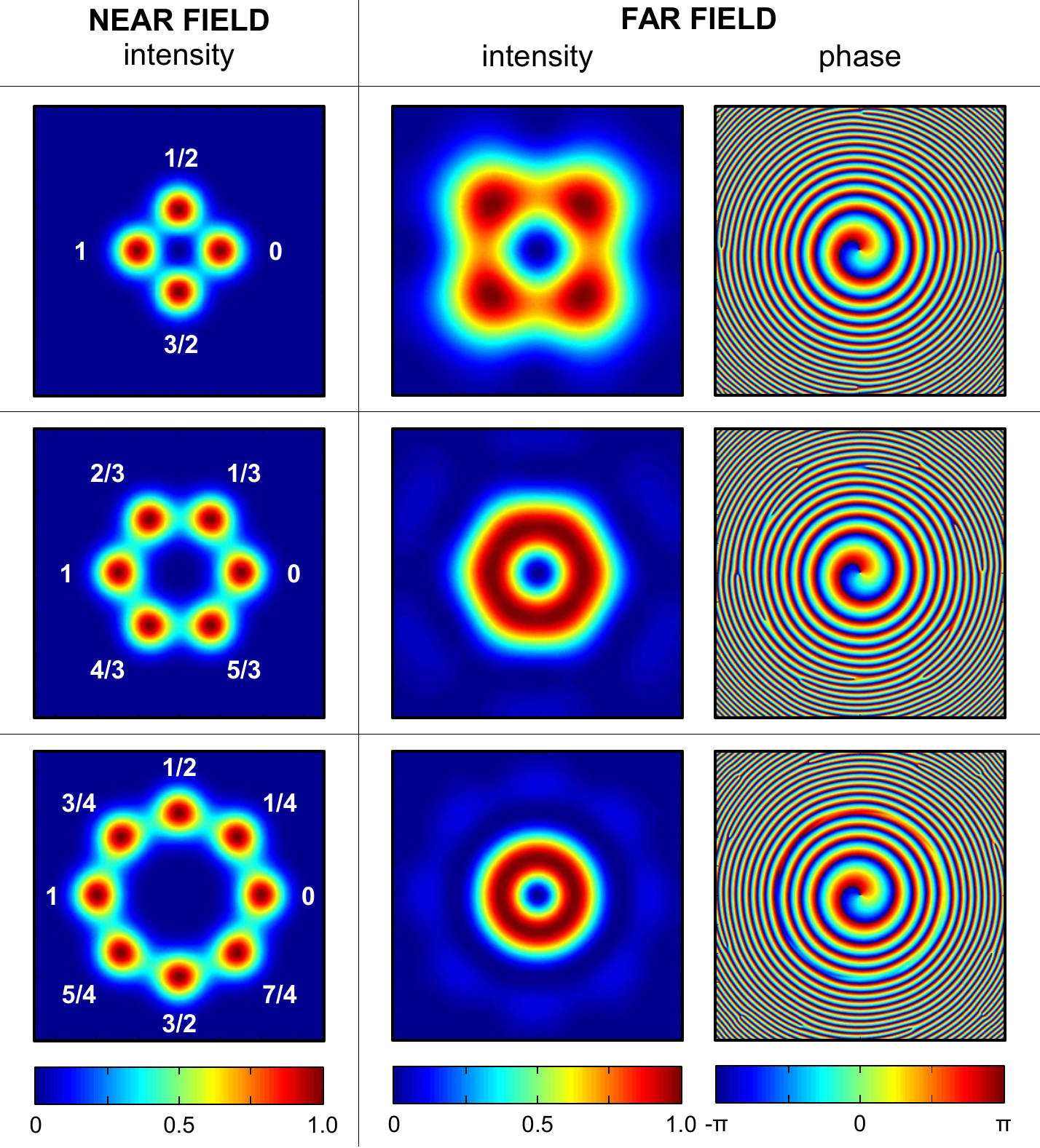}
\vspace{1.5cm}
\caption{Examples of near-field and far-field distributions at the output of waveguide arrays, where the waveguides are placed on the vertices of a regular polygon and modes are considered as Gaussian with diameter $2 w_0$~=~15~$\mu$m (and wavelength $\lambda$~=~633~nm). For the near field the normalized intensity distribution is shown, and the phase delay of the field of each waveguide is indicated, in multiples of $\pi$. For the far field the normalized intensity and the phase of the electromagnetic field are reported in distinct graphs, calculated at a distance $z$~=~15~mm from the output of the array. Picture size is 70~$\mu$m~$\times$70~$\mu$m for the near field pictures and 1~mm~$\times$~1~mm for the far field ones.}
\label{fig:DifferentPolygons}
\end{figure*}

Three-dimensional waveguide arrays have been also recently investigated to this purpose. Markin and coauthors proposed to directly excite supermodes of a triangular array of coupled waveguides, having cyclical phase distribution, by means of a phase-matched nonlinear process \cite{markin2013}. Guan et al. showed the possibility to encode 15 different states with cyclical phase distribution in a set of 16 non-coupled waveguides, placed on the vertices of a regular polygon: this was achieved by splitting the light in the different waveguides with an integrated-optics star coupler and by using thermo-optic phase shifters for fine adjustment of the phase terms \cite{guan2014}. However, these previous works were limited to the study of the propagating field within the array or to the near field distribution, while the far-field distribution was not studied or employed.

As a matter of fact, one could show that a single linearly-polarized mode or supermode,or a set of separated linearly-polarized waveguide modes has in each point a transverse component of the Poynting vector with vanishing time average (see \ref{appendixVanishingOAM}). This prevents light propagating on that mode or those modes to have a physical, non-vanishing, longitudinal angular momentum within the array, even in limited regions. However, this does not hold for overlapped optical modes in free space propagation, and thus stimulates the investigation of the free-space beams emitted at the output of this kind of waveguide arrays.

Namely, we study the far-field distribution produced by the output of an array of $N$ waveguides, placed (in the cross-section) at the vertices of a regular polygon, with a cyclical phase distribution, similar to the one demonstrated in Ref.~\cite{guan2014}. 
As discussed in \sref{sec:general}, the near field of the output of such an array can be described with good approximation as a set of Gaussian optical modes, that we place in the positions:
\begin{equation}
\bi{R}_n = R_0 \cos(2 \pi \frac{n}{N}) \bi{u}_x +  R_0 \sin (2 \pi \frac{n}{N}) \bi{u}_y
\end{equation}
where $R_0$ is the radius of the polygon, and we consider the phases of the different modes to be arranged as:
\begin{equation}
\alpha_n = 2\pi \frac{n}{N}  l
\label{eq:phases}
\end{equation}

\Fref{fig:DifferentPolygons} shows some examples of near-field and far-field distributions for some of the simplest cases in which $l$~=~1 (phases make a 2$\pi$ round-trip around the polygon), and polygons with $N$~=~4,~6,~8. We consider mode size ($2~w_0$~=~15~$\mu$m) and wavelength ($\lambda$~=~633~nm) commonly employed in three-dimensional waveguide arrays fabricated by femtosecond laser writing \cite{szameit2006}. The far-field distribution is calculated by directly evaluating \eref{eq:campoArrayGeneral} for $z$~=~15~mm. The presence of a phase singularity in the center of the far-field distribution is easily observed in all cases, associated with an intensity distribution that closely resembles the typical doughnut shape of the Laguerre-Gauss modes, i.e. the archetypical examples of beams with non-vanishing angular momentum \cite{allen1992}.

\begin{figure*}[tbp]
\centering
\includegraphics[width=\textwidth]{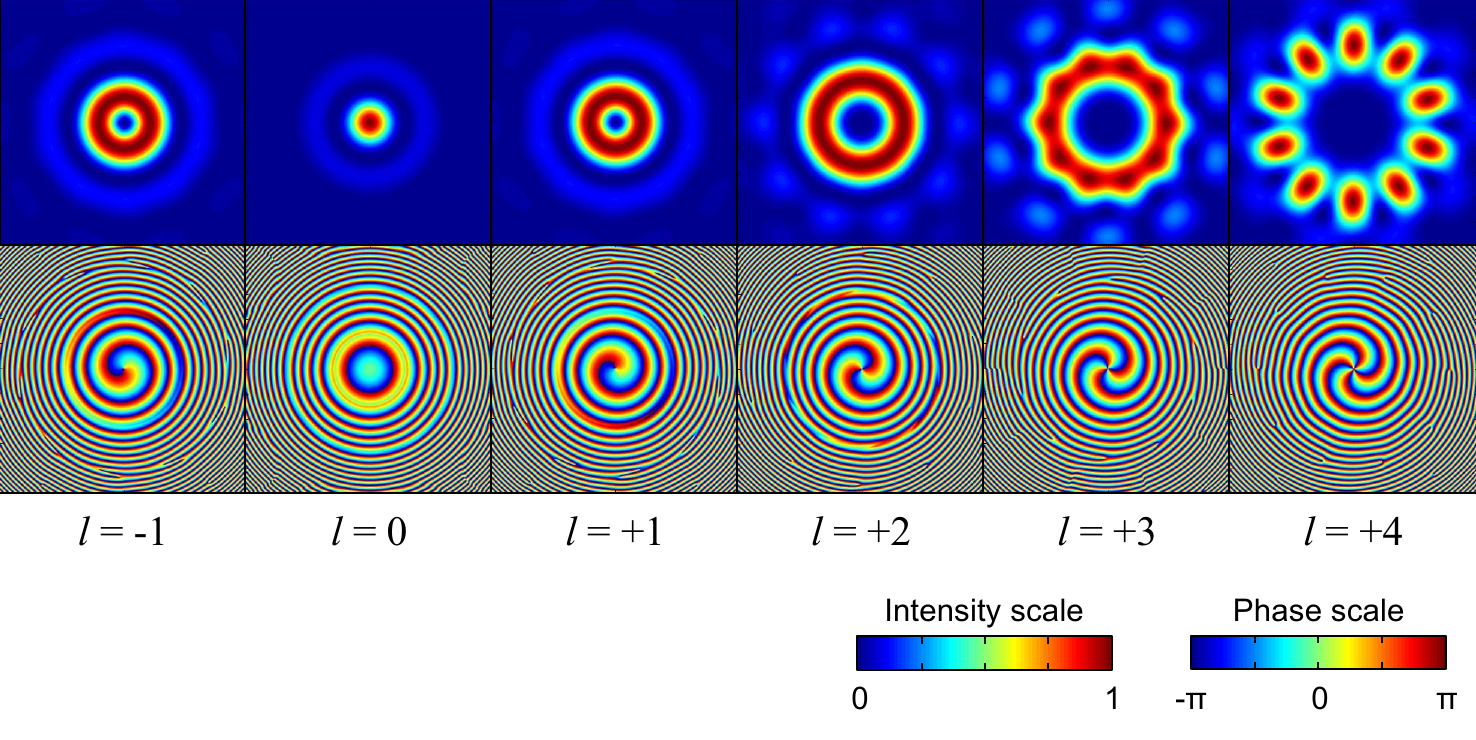}
\caption{Examples of far-field distributions at the output of waveguide arrays, where the waveguides are placed on the vertices of a regular polygon with $N$~=~10 edges, with radius $R_0$~=~25~$\mu$m. Modes are considered as gaussian with diameter $2 w_0$~=~15~$\mu$m (and wavelength $\lambda$~=~633~nm). Phase delays are arranged in the different waveguides according to $\alpha_n = 2\pi \frac{n}{N}  l$, with different values of $l$. Graphs of the normalized intensity (top pictures) and the phase of the electromagnetic field (lower pictures) are reported, calculated at a distance $z$~=~15~mm from the output of the array. Each picture represents an area of 1~mm~$\times$~1~mm.}
\label{fig:DifferentAngularMomentum}
\end{figure*}
The presence of a phase singularity in the center of these field distributions is shown analytically in \ref{appendixLaguerreGauss}; in particular, it is shown that \eref{eq:campoArrayGeneral}, which describes the field of the array, converges to
\begin{equation}
u(r, \phi, z) \propto r^{|l|}   e^{\iota  l \phi} 
\end{equation}
for small $r$ and for large $z$ ($r$, $\phi$ and $z$ are the radial, azimuthal and longitudinal coordinates respectively). Indeed, this same expression can be found as a limit for small $r$ in the case of Laguerre-Gauss beams \cite{allen1992} or Bessel beams \cite{mcgloin2005} of order $l$, which are well-known examples of beams carrying orbital angular momentum. Thus, this ensures that the beam is carrying orbital angular momentum proportional to $l$ in its central region.

In fact, field distributions with different angular momentum values can be produced as shown in \fref{fig:DifferentAngularMomentum}, which reports the far-field distribution of a polygonal waveguide array with $N$ = 10 waveguides, and phases arranged cyclically for different values of $l$ in \eref{eq:phases}. Different phase vortices can be realized with both clockwise and counter-clockwise orientation, depending on the sign of $l$. On the other hand, for $l$~=~0 the far-field distribution converges to a pattern with a single peak in the middle. The doughnut shape of the intensity distribution is less defined for increasing $|l|$; in fact, the doughnut would extend farther from the center, where the approximations of \ref{appendixLaguerreGauss} become less accurate. We can further observe that the periodicity of the phase cycle needs not to be commensurate with the number of waveguides composing the array; for instance it is possible in our case to achieve $l$~=~3. Finally, we should of course note that it is not possible to effectively produce vortices with $l$ equal or higher than $N/2$: from the sampling theorem it is clear that in $N$ distinct phase values (as available in $N$ waveguides) there would not be enough information to produce such vortices.

\section{Towards more complex field configurations}
\label{sec:complex}

We shall now discuss the potentials of our approach in producing more complex far-field distributions.

\begin{figure*}[tbp]
\centering
\includegraphics[width=\textwidth]{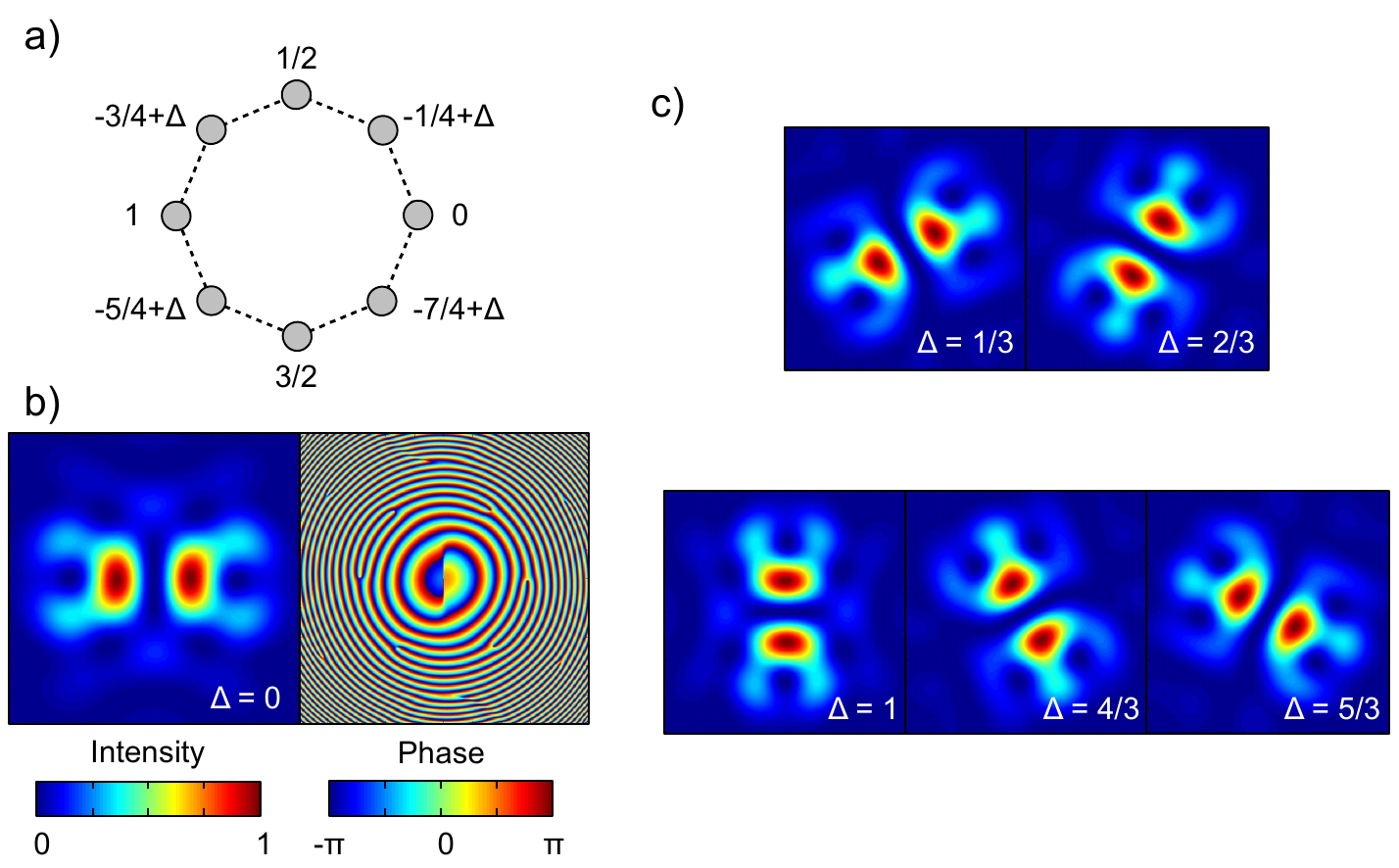}
\caption{Study of the far-field distribution of an octagonal waveguide array with radius $R_0$~=~20~$\mu$m, where the phases of the different optical modes at the output are arranged as shown in panel \textit{(a)} in multiples of $\pi$. Modes are considered as Gaussian with diameter $2 w_0$~=~15~$\mu$m (and wavelength $\lambda$~=~633~nm). \textit{(b)}~Intensity and phase distribution of the field, calculated at a distance $z$~=~15~mm, for $\Delta$~=~0. \textit{(c)}~Intensity distributions for different other values of $\Delta$. All pictures have the same scale, and correspond to a region 1~mm~$\times$~1~mm wide.}
\label{fig:HermiteGauss}
\end{figure*}
The first case we consider is illustrated in \fref{fig:HermiteGauss}. Eight waveguides are placed on the vertices of an octagon, and are divided into two series (say, the odd waveguides and the even waveguides). All waveguide modes carry the same intensity. The odd modes have cyclic phases arranged in counter-clockwise direction, while the even modes have cyclic phases arranged in clockwise direction. A phase shift $\Delta$ is further added to the phases of the even modes. This produces the two-lobes far-field distributions, whose orientation depends on $\Delta$, that can be seen in \fref{fig:HermiteGauss}  (b) and (c).

If only the odd or even waveguides were present, we would observe in the far-field a vortex beam distribution such as those in \fref{fig:DifferentPolygons}, with an asymptotic similarity (in the central part) to the Laguerre-Gauss modes with $l$~=~$\pm$1, as shown in \ref{appendixLaguerreGauss}. However, here we have the linear combination of two vortices with opposite direction. It is known that the equally weighted combination of two Laguerre-Gauss modes with $l$~=~$\pm$1 produces a two-lobe Hermite-Gauss mode, oriented along an axis which depends on the phase delay between the two Laguerre-Gauss modes. Indeed, we observe an analogous behaviour in \fref{fig:HermiteGauss}. 

We find this example interesting because it shows the possibility to change profoundly both the phase and the intensity profile of the far-field just by acting on the phase terms of the modes in the array, starting from the same balanced intensity distribution. In fact, from the same eight waveguides arranged on the vertices of an octagon, one could produce phase vortices with different orientation or vorticity such as the ones in \fref{fig:DifferentAngularMomentum}, or other configurations with zero angular momentum such as the ones in \fref{fig:HermiteGauss}, just by acting on those phases. In waveguide devices, one may think of accurately controlling such phase terms even in a dynamic fashion, by e.g. thermo-optic phase shifters \cite{guan2014, flamini2015}.

To guide the design process of waveguide arrays that produce phase and intensity patterns in the far-field with even richer features, it can be useful to go back to a more general picture of the involved diffraction phenomena. In fact, in the Fraunhofer diffraction condition, i.e. for sufficiently large $z$, it is well known that the far-field can be approximated by the Fourier transform of the near-field \cite{goodman2005}. More precisely, by dropping common phase terms, one can write in our case:
\begin{equation}
u_f (\bi{r},z) \simeq e^{\iota \frac{\pi}{\lambda z} |\bi{r}|^2 } \, \cdot \, \mathcal{F} \left \lbrace g(\bi{r},0) \right \rbrace \cdot \mathcal{F} \left \lbrace \sum^N_{n=1} \sqrt{I_n} e^{\iota \alpha_n} \delta(\bi{r} - \bi{R}_n) \right \rbrace
\label{eq:campoTrasformato}
\end{equation}
where the transform $\mathcal{F}$ is defined as:
\begin{equation}
\mathcal{F} \lbrace f \rbrace = \int \!\!\! \int^{+\infty}_{-\infty} f(x',y')
e^{- \iota\frac{2\pi}{\lambda z}  \left( x x' + y y' \right)} dx' dy'
\end{equation}
and indeed consists in a properly scaled two-dimensional Fourier transform. Practically, the far-field distribution in our case is given by an interference pattern, which is the Fourier transform of the position of the modes, modulated by the diffraction pattern of a single waveguide mode $\mathcal{F} \left \lbrace g(\bi{r},0) \right \rbrace$. 

One can think, as a consequence, to devise the structure of a waveguide array by, first, anti-transforming a certain desired far-field distribution, and then trying to approximate with a pattern of Gaussian modes (with identical shape but different phases or intensities) the result of this anti-transformation. Naturally, knowledge of the fundamental properties of the Fourier transform is of great help. In particular, it may be useful to remind that the transform of a regular lattice of peaks is again a regular lattice (the reciprocal lattice). Far-field configurations showing regular lattice structure may thus be easier to reproduce by means of the far-field emission of a waveguide array (with a certain amount of regularity).

\begin{figure*}[tpb]
\centering
\includegraphics[width=0.8\textwidth]{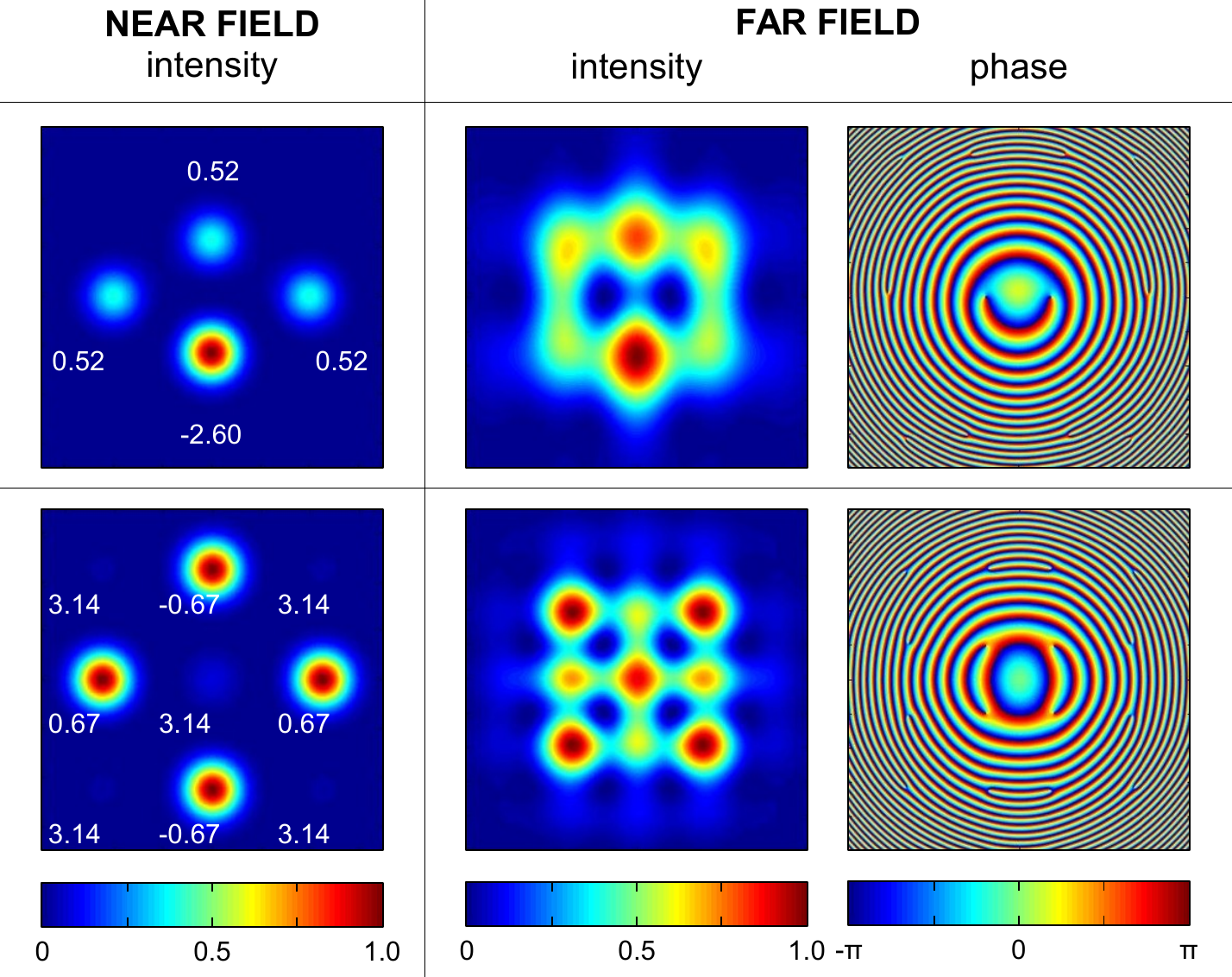}
\caption{Examples of non-trivial field configurations that can be achieved with the output of waveguide arrays with proper waveguide disposition and relative phase. The column on the left shows the near-field distributions. Each Gaussian mode has a diameter $2 w_0$~=~15~$\mu$m and wavelength $\lambda$~=~633~nm. The phases of the different modes are indicated in radians near each mode in the pictures. In the right columns the corresponding intensity and phase distributions in the far-field are shown, calculated at a distance $z$~=~15~mm. We can observe the presence of two (example on the top row) and four (example on the bottom row) main points of phase singularity, which also coincide with zeroes in the intensity distribution. Near-field pictures show an area of 70~$\mu$m~$\times$~70~$\mu$m, while the far-field ones show an area of 1~mm~$\times$~1~mm.}
\label{fig:ComplexDistr}
\end{figure*}

\Fref{fig:ComplexDistr} shows two peculiar examples, that have been engineered following the above described procedure, i.e. by approximating with the waveguide position, intensity and phases, the anti-transform of a certain desired far-field. The first one (pictures on the top row) was designed looking for an ``eyeglass'' shape, with two counter-rotating and neighbouring phase vortices. The starting point to devise the second one (pictures on the bottom row) was a 2$\times$2 matrix of phase singularities, two encircled by a clockwise phase orientation, two encircled by phases rotating in the opposite direction. It can be observed that the achieved far field configurations, that show two or four main points of phase singularity, can be produced by very simple dispositions of few Gaussian modes, provided that the relative phases and intensities are set with sufficient accuracy.
It may also worth noting that the spiralling phase structure, which is easily observed around the phase singularity points in figures \ref{fig:DifferentPolygons} or \ref{fig:DifferentAngularMomentum} here is not apparent, because it is dominated by the radial phase variation due to the curvature of the wavefront.

\section{Propagation of the projected beam}
\label{sec:propagation}

Having shown the possibility to engineer the intensity and phase distributions in the far-field, it can be interesting to analyse the intensity pattern of the projected beams during their propagation. In fact, as we will show, by properly designing the waveguide position and phases it is possible to influence relevant properties of the propagating beam, such as the focus position.

\begin{figure*}[tpb]
\centering
\includegraphics[width=0.9\textwidth]{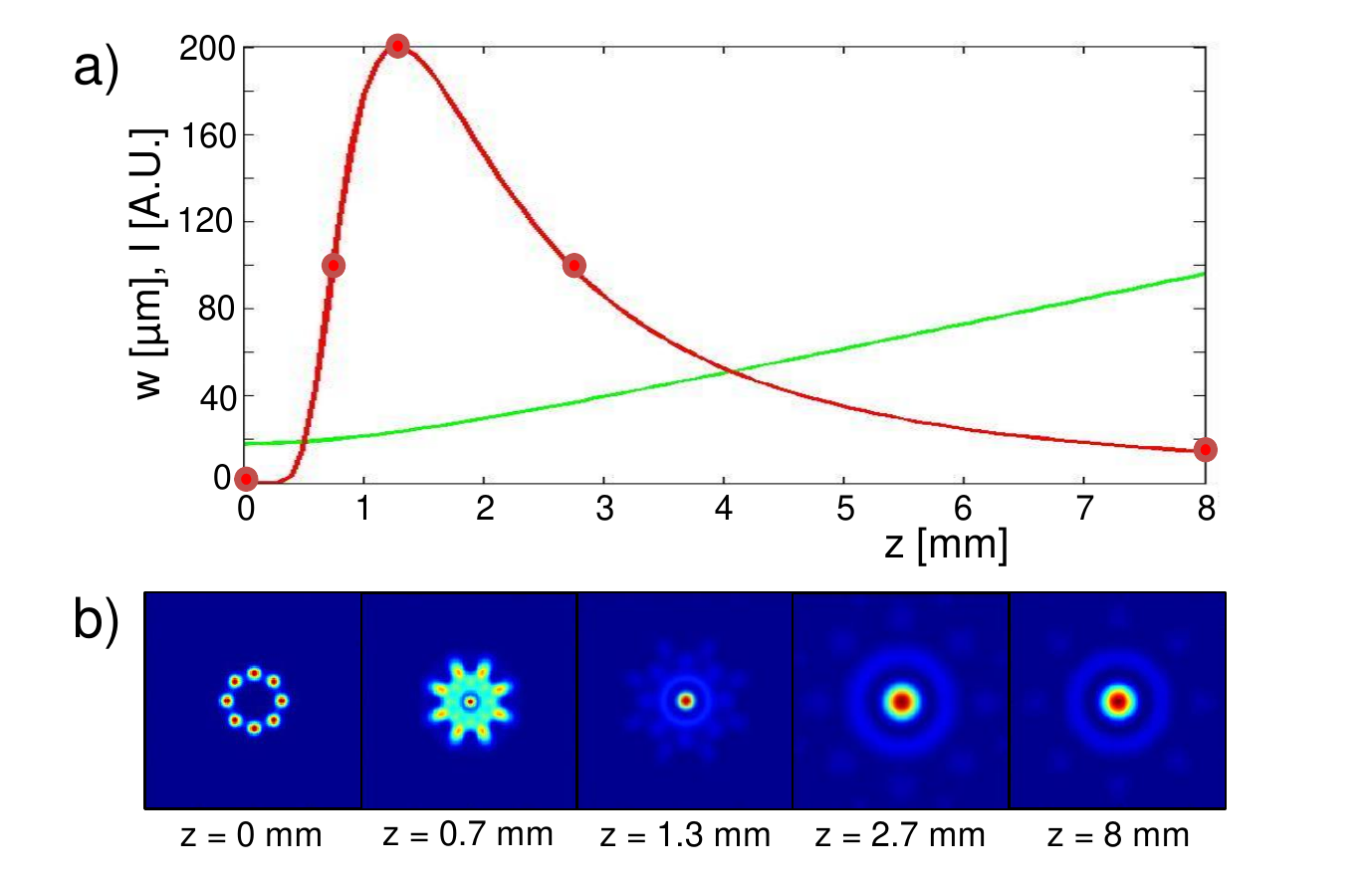}
\caption{Evolution of the intensity pattern during the propagation of the beams output of a waveguide array where the waveguides are placed on the vertices of a regular polygon with $N$~=~8 edges, with radius $R_0$~=~25~$\mu$m. Modes are considered as Gaussian with diameter $2 w_0$~=~15~$\mu$m (and wavelength $\lambda$~=~633~nm). \textit{(a)} Intensity profile along the propagation axis and the spot size of the beam during the propagation. \textit{(b)} Spatial intensity distribution at the propagation distances indicated with circles in panel (a), {\it i.e.} in the near-field and far-field, and at the distances where the axial intensity is equal to the maximum and half of it. Transverse intensity pictures show an area of 200~$\mu$m~$\times$~200~$\mu$m, except for the far-field one that show an area of 600~$\mu$m~$\times$~600~$\mu$m.}
\label{fig:Bessel_evolution}
\end{figure*}

We can start our analysis from the case of the polygonal arrays, already discussed in \sref{sec:polygonal} for what concerns the far-field. As already observed (see \fref{fig:DifferentAngularMomentum}), if the waveguide phases are all equal to each other, the projected beam  shows in the far-field an intensity pattern with a single peak in the middle, surrounded by a less intense ring, a distribution that is similar to that of a zero-order Bessel beam. A Bessel beam is usually realized by illuminating an axicon with a Gaussian beam or by placing an annular aperture in the back focal plane of a focusing lens \cite{mcgloin2005}. A similar configuration can be easily achieved through the use of few waveguides positioned on the vertices of a polygon.  A field distribution with a spatial profile approaching that of a Bessel beam can indeed be obtained, as shown in \fref{fig:Bessel_evolution}, by exploiting an array with  $N$ = 8 waveguides placed on the vertices of a regular octagon of radius $R_0$~=~25~$\mu$m. Bessel beams are ideally supposed to be non-diffracting beams, a peculiar property that can be of interest in many application. However in practical cases they show this behaviour only in a limited region, the so-called Bessel zone, which can be estimated by calculating the propagation length where the intensity of the central peak is higher or equal to half of the maximum intensity.  

In \fref{fig:Bessel_evolution} (a) we report the evolution of the intensity of the central peak along the propagation direction. From the analysis of the curve we can estimate the length of the Bessel zone, which is approximately 2~mm long, {\it i.e.} much longer than the Rayleigh range of a Gaussian beam with the same wavelength and waist as the waveguides of the array, which is about $z_R$ = 0.28~mm. In \fref{fig:Bessel_evolution} (b) we also report the transverse intensity pattern at specific propagation distances, that is, in the near-field at $z$ = 0~mm, at the coordinates where the axial intensity is equal to the maximum value, to its half and finally at a large distance, which can be considered in the far-field. It can be observed that in the first part of the propagation a reshaping of the intensity pattern of the array occurs, reaching a Bessel-like distribution with a central peak surrounded by rings where the intensity is maximum. Afterwards, the beam conserves its shape during the propagation, showing only a diffraction of the pattern.
We calculated also the spot size parameter of the beam \cite{svelto4th} at various distances, defined as $w_{x,y}(z) = 2 \sigma_{x,y}(z)$ where $\sigma_{x,y}^2$ is evaluated along the $x$ or $y$ coordinate as: 
\begin{equation}
\fl
\sigma^2_x (z) = \frac{\int \!\!\! \int (x - \langle x \rangle)^2 I(x,y,z) dx dy}{\int \!\!\! \int I(x,y,z) dx dy}  \qquad \sigma^2_y (z) = \frac{\int \!\!\! \int (y - \langle y \rangle)^2 I(x,y,z) dx dy}{\int \!\!\! \int I(x,y,z) dx dy}  
\label{eq:spotsize}
\end{equation}
The result is shown in \fref{fig:Bessel_evolution}, and confirms the beam divergence during the propagation. We can thus  conclude that the near-field configuration considered in this example experiences a reshaping of the spatial intensity pattern with a sort of focusing after $\sim$1.3 mm of propagation.

\begin{figure*}[tpb]
\centering
\includegraphics[width=\textwidth]{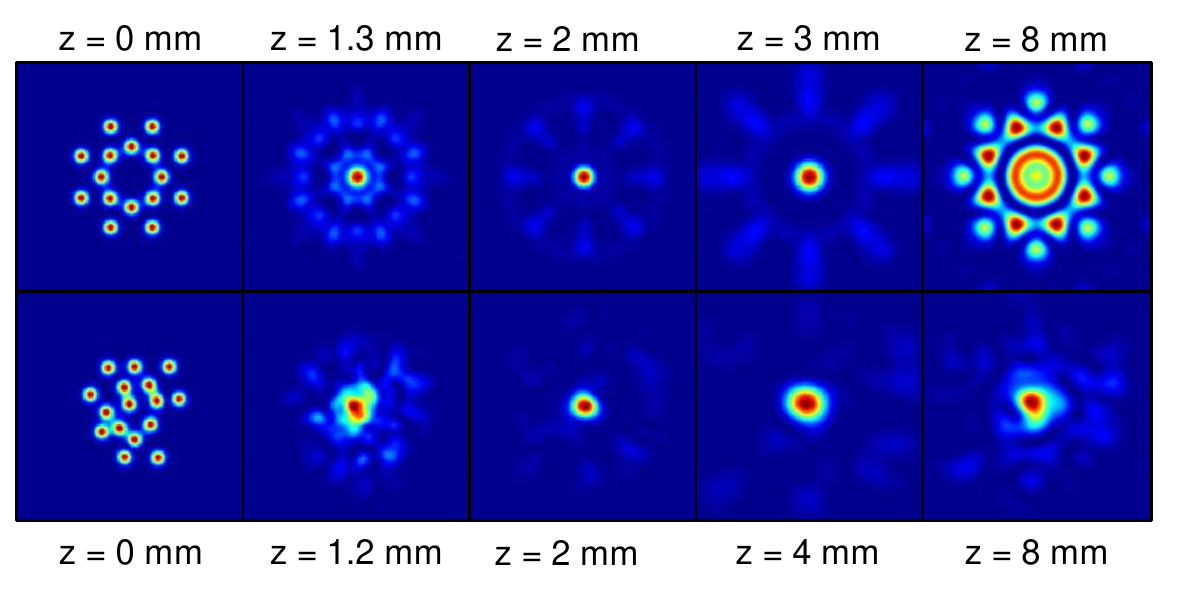}
\caption{Evolution of the intensity pattern of the beams emitted by a waveguide array where the waveguide phases are adjusted according to \eref{eq:fasiFresnel} so as to move the focusing position to $z$~$\simeq$~2~mm. Top row: the waveguides are placed on the vertices of two regular polygons with $N$~=~8 edges, respectively with radii $R_0$~=~25~$\mu$m and $R_0$~=~45~$\mu$m. Bottom row: the waveguides are distributed in random positions.  Modes are considered as Gaussian with diameter $2 w_0$~=~15~$\mu$m (and wavelength $\lambda$~=~633~nm). Transverse intensity pictures show an area of 200~$\mu$m~$\times$~200~$\mu$m, except for the far-field ones that show an area of 600~$\mu$m~$\times$~600~$\mu$m.}
\label{fig:focusing}
\end{figure*}

It is well known in diffractive optics that, by governing the phase front in the near field, it is possible to control the intensity pattern along the propagation, e.g. to achieve focusing or to engineer the shape of the focus itself. As an example, phase shaping has also been used to make the peak intensity evolution of a Bessel-gauss beam uniform within the Bessel zone \cite{Moloney2013}.

Here, we can think of moving the focus position by properly adjusting the position and phases of the waveguides in polygonal arrays, in particular considering the propagation of the field emitted by waveguides positioned on annuli with different radii and phase shifts. This configuration can be assimilated to that of a Fresnel lens, which is made of a series of concentric rings with different radii, each designed to diffract light so as to obtain constructive interference in the position where the focus is desired. This condition can be achieved by adjusting the phases of the waveguides in each ring according to the following equation:
\begin{equation}
\alpha_n = \frac{2 \pi}{\lambda} \sqrt{r_n^2+f^2} 
\label{eq:fasiFresnel}
\end{equation}
where $f$ is the desired focal distance and $r_n= \sqrt{x_n^2+y_n^2}$ the radial coordinate of the waveguide position.
As an example, we consider a near-field pattern where the waveguides are distributed over two concentric rings with radii $R_1$~=~25~$\mu$m and $R_2$~=~45~$\mu$m. By adjusting the phases in such a way that the condition of constructive interference is satisfied for a focal distance of 2~mm, we obtain the translation of the propagation distance at which the intensity is maximum from $z$~=~1.3 mm, as obtained with only one ring, to $z$~=~2 mm. The results are shown in the top row of \fref{fig:focusing}, where we report the transverse intensity pattern at different propagation distances. 

Finally it is worth noting that, if one can separately act on the phase terms of a large number of waveguides, an efficient focusing can be obtained also when starting from a random distribution of the waveguides in the array, provided that the phases are properly set by the relation \eref{eq:fasiFresnel}. An example of the intensity pattern  that can be achieved at different propagation distances is reported in \fref{fig:focusing}, where, starting from a near-field configuration of 16 waveguides randomly distributed in a 50~$\mu$m radius circular area, a focal zone is obtained after 2~mm of propagation. To evaluate the focusing efficiency we calculated the spot dimension at 1/e$^2$ and the ratio between the fraction of power confined in this region for the configuration in analysis, and for a Gaussian beam. The random configuration shown in \fref{fig:focusing} yields a focusing efficiency of 50\%, higher than the efficiency of 28\% that we achieved in the two-rings case.

\section{Feasibility considerations and conclusions}
\label{sec:conclusion}

Before concluding, we should consider whether 
the devices and waveguide configurations presented in the previous sections can be effectively fabricated, or our essay has to be considered a mere speculation. The three-dimensional capabilities of femtosecond laser writing technology certainly enables to produce arrays of waveguides with arbitrary positions in the cross-section \cite{szameit2006,rechtsman2012,caruso2016}. However, it may be more difficult to divide the input light, e.g. from an optical fiber, into the waveguides of the array with the required different amplitude ratios and phases.

To this purpose, a very general scheme may be a generalisation of the one adopted in Ref.~\cite{guan2014}, where the integrated circuit is divided into three sections. The first section consists in a power dividing device, such as a star coupler or a cascade of directional couplers, to split the input light from the fiber into different branches with the required amplitude ratios. The second section consists in an array of phase shifters (e.g. thermo-optic devices) that allow to tune the phase terms of each waveguide. These sections may be fabricated in the plane and possibly with lithographic technologies. A third section transforms the waveguides from a planar arrangement to the required two-dimensional cross-section, bringing each waveguide to its final position. This needs to be realized by femtosecond laser writing due to the three-dimensionality of the layout. To minimise unwanted parasitic coupling between the waveguides, in the region where they are brought close together to produce the final setting, one might fabricate neighbouring waveguides with sufficiently different propagation constants, so that evanescent field coupling is quenched.
With a higher technological effort, the first two sections may be integrated in a single fully-reconfigurable multiport interferometer that enables realising arbitrary unitary transformations between the input and the output waveguides \cite{carolan2015}. This would also permit to associate to each different input waveguide a different output configurations, provided that the chosen alternative output configurations are orthogonal supermodes.

As a matter of fact, the capabilities of femtosecond laser waveguide writing could allow reducing the three sections to a single three-dimensional interferometer, at least for specific applications. Let us consider, as an example, a possible way to produce different vortex configurations, as in \fref{fig:DifferentAngularMomentum}, with the same physical array of $N$ waveguides. The different vortex configurations have all the waveguides excited with the same amplitude; the phase terms however are arranged cyclically, with different periodicity. These field configurations may be produced with a $N$-inputs-$N$-outputs device that implements the Fourier transform of the input modes \cite{barak2007}: by exciting a different input of the device, a different phase arrangement of the output would be produced. Compact integrated optical circuits performing this operation have been indeed recently demonstrated experimentally \cite{crespi2016}.

In the light of these considerations, our approach looks reasonably at reach of current technology. In particular,  we have shown that it is possible to design waveguide arrays emitting at the output light beams with one or more phase singularity points, possibly carrying non-zero angular momentum. Focusing beams can also be produced, by adequately controlling the phase terms. In addition, the investigation of the far-field distribution here presented is immediately applicable to the distribution in the focal plane of a lens. 

This work paves the way to other investigations that could deepen many aspects here just mentioned. For instance, for free-space communication purposes it will be required to understand how to optimize the array design to minimize cross-talks between the emitted modes \cite{zhao2015}. The capability to build multi-mode interferometric circuits that associate to each different input a different emitted supermode (e.g. with a different value of angular momentum), could be conveniently used as angular momentum or space division multiplexer/demultiplexer from a fiber optics networks \cite{su2012, cai2012,guan2014}. In quantum optics applications, this would also allow to transfer entanglement from the waveguide path to the angular momentum or other spatial degree of freedom of the photons \cite{fickler2014}. 

Benefits are also envisaged in optofluidics where lab-on-a-chips are used for single cell manipulation through optical forces \cite{yang2016}. 
Indeed the integration of engineered waveguide arrays in optofluidic lab-on-a-chips could allow to implement tweezing or manipulation functionalities, as the control of the particle rotation in the optical trap for imaging and analysis purposes, which nowadays require bulk optics equipment. Unexpected applications might also be found in beam shaping for structured illumination, optical filtering or imaging tasks.

\appendix

\section{Vanishing angular momentum within the array}
\label{appendixVanishingOAM}
We will show here that light propagating in a set of linearly-polarized and spatially separated guided optical modes cannot carry longitudinal angular momentum. The supporting dielectric structure and refractive index distribution will be considered as uniform along the $z$ axis, so that such axis is a well defined propagation direction and the optical mode profiles extend in the $xy$ plane.

The propagating electromagnetic field of each mode, in each point, decomposed into its transverse and longitudinal components, takes the form:
\begin{equation}
	\begin{array}{ll}
		\bi{E}_{\perp} = \bi{E}^0_{\perp} (x,y) \, e^{\iota (\omega t - kz)} &
		\bi{E}_{z} = \iota E_{z}^0 (x,y) \, e^{\iota (\omega t - kz)} \bi{u}_z\\
		\bi{B}_{\perp} = \bi{B}_{\perp}^0 (x,y) \, e^{\iota (\omega t - kz)} &
		\bi{B}_{z} = \iota B_{z}^0 (x,y) \, e^{\iota (\omega t - kz)} \bi{u}_z
	\end{array}
\end{equation}
The transverse Poynting vector, responsible for the longitudinal angular momentum, can be written as:
\begin{equation}
\bi{S}_\perp = \frac{1}{2\mu_0} \left( \bi{E}_\perp \times \bi{B}_z + \bi{E}_z \times \bi{B}_\perp \right)
\end{equation}
and its time average is:
\begin{eqnarray}
 \bi{S}_\perp^\mathrm{ave} = \frac{1}{2 \mu_0} \left( \iota E_z \bi{u}_z \times \bi{B}_\perp^* + \bi{E}_\perp \times (-\iota B_z \bi{u}_z) + \right. \nonumber \\ 
\left. + (- \iota E_z \bi{u}_z) \times \bi{B}_\perp + \bi{E}^*_\perp \times \iota B_z \bi{u}_z \right)
\end{eqnarray}

If $\bi{E}_\perp$ and $\bi{B}_\perp$ are taken as real, it naturally holds that
 $\bi{E}_\perp = \bi{E}^*_\perp$ and  $\bi{B}_\perp = \bi{B}^*_\perp$. It is then easy to observe that, in each point $\bi{S}_\perp^\mathrm{ave} = 0$.
Since the transverse component of the angular momentum density, with respect to an arbitrary point O, is $\bi{l}_z = \varepsilon_0 \mu_0  \left( \bi{r} - \bi{r}_O \right) \times \bi{S}_\perp$, in each point of the mode field its time average will be vanishing:
\begin{equation}
\bi{l}^\mathrm{ave}_{z,O} = \varepsilon_0 \mu_0  \left( \bi{r} - \bi{r}_O \right) \times \bi{S}^\mathrm{ave}_\perp = 0 \label{eq:amDensityAve}
\end{equation}
The overall angular momentum of the $n$-th mode field would be $\bi{L}^n_z = \int \bi{l}^n_z d\tau$, where the integration is performed over all the volume of the space and only the contribution of the $n$-th waveguide field has been considered. It is clear  from \eref{eq:amDensityAve} that also the time average of the overall angular momentum is vanishing.

If we consider a set of well-separated modes, the overall angular momentum of the field is just $\bi{L}_z = \sum^N \bi{L}^n_z$, which has also vanishing time-average. The field of an array of well separated, linearly polarized, optical guided modes, thus possesses no angular momentum. Excitation of several overlapped modes is a necessary condition to have angular momentum within the array: in fact, in that case, since the modes are overlapped, the Poynting vector of the resulting field is not in general equal to the sum of the Poynting vector of the modes.

\section{Phase singularity in the far-field of a polygonal waveguide array}
\label{appendixLaguerreGauss}

We derive here an asymptotic expression for the center of the far-field distribution of a polygonal waveguide array, as those described in \sref{sec:polygonal}, valid in the limits of large propagation coordinate $z$ and small distances $r$ from the center.

We consider $N$ Gaussian modes, monochromatic and linearly polarized, all having their waist at $z$ = 0 coordinate, and centered on the vertices of a regular polygon:
${\bi R}_n = R_0 \cos(2 \pi \frac{n}{N}) {\bi u}_x +  R_0 \sin (2 \pi \frac{n}{N}) {\bi u}_y$. All the Gaussian modes have the same intensity, but different phase terms, arranged in a cyclical configuration $\alpha_n = 2\pi \frac{n}{N}  l$. The scalar field, at a certain plane $z$, has the form :
\begin{equation}
u = u_0(z) \sum_{n=1}^N e^{+\iota \alpha_n} e^{-\frac{\left| \bi{r} - \bi{R}_n \right|^2}{w(z)^2}} e^{-\iota k \frac{ \left| \bi{r} - \bi{R}_n \right|^2}{2 \rho(z)}} \label{eq:campo1}
\end{equation}
where $\bi{r}= r \cos(\phi) \bi{u}_x + r \sin(\phi) \bi{u}_y$, $k = 2 \pi / \lambda$, $w(z) =\sqrt{w_0^2 + \frac{\lambda^2 z^2}{\pi^2 w_0^2}}$ and $\rho(z)$ is the curvature radius of the wavefront. It can be assumed $\rho(z) \simeq z$ for $z$ sufficiently large. $u_0(z)$ is a normalization factor which includes common $z$-dependent phase terms.

We start by expanding the squares in the arguments of the exponentials in \eref{eq:campo1}, as $\left| \bi{r} - \bi{R}_n \right| = r^2 + R^2_0 - 2 r R_0 \cos \left( 2 \pi \frac{n}{N} - \phi \right)$, so that:
\begin{eqnarray}
\fl
u(r,\phi,z)  = \left( u_0(z) e^{-\frac{R_0^2}{w(z)^2} + \frac{\iota k R_0^2}{\rho(z)} } \right) e^{-\frac{r^2}{w(z)^2}} e^{-\iota k \frac{r^2}{2 \rho(z)}}  \sum_{n=1}^N e^{-\iota \alpha_n} e^{\frac{ 2 r R_0 \cos \left( 2 \pi \frac{n}{N} - \phi \right) }{w(z)^2} + \iota k \frac{ 2 r R_0 \cos \left( 2 \pi \frac{n}{N} - \phi \right)}{2 \rho(z)}} \nonumber \\
\end{eqnarray}
In the far field, i.e. for $z$ sufficiently large, $\frac{\rho(z)}{k} \simeq \frac{z}{k} \ll w(z)^2$, thus:
\begin{equation}
u(r,\phi,z)  \simeq u'_0(z) e^{-\frac{r^2}{w(z)^2}} e^{-\iota k \frac{r^2}{2 z}}  \sum_{n=1}^N e^{\iota 2\pi \frac{n}{N}  l} e^{\iota \frac{k}{2 z} 2 r R_0 \cos \left( 2 \pi \frac{n}{N} - \phi \right) }
\label{eq:AppB_passage1}
\end{equation}
where all the constant terms have been included in $u'(z)$.

We shall now focus on the central part of the beam and we seek for a power series limit of \eref{eq:AppB_passage1}, valid for small $r$. In particular, we consider:
\begin{equation*}
r \ll \frac{z}{k R_0} = \frac{z \lambda}{2 \pi R_0}
\end{equation*}
We note that in the same limit, since $w(z) \simeq \frac{\lambda z}{\pi w_0}$ we have also (and with much better approximation) $r \ll w(z)$.

This allows to substitute the last exponential in the sum with its power series expansion, and to approximate $e^{-\frac{r^2}{w(z)^2}}\simeq 1$, i.e.:
\begin{eqnarray}
\fl
u(r,\phi, z)  \simeq u'_0(z) e^{-\iota k \frac{r^2}{2 z}} \sum_{n=1}^N e^{\iota 2\pi \frac{n}{N}  l} \sum_{m=0}^\infty \frac{ \iota^m }{m!} \left( \frac{k}{ z} r R_0 \right)^m \cos^m (2\pi  \frac{n}{N} - \phi) = \nonumber \\
= u'_0(z) e^{-\iota k \frac{r^2}{2 z}} \sum_{m=0}^\infty  \frac{ \iota^m }{m!} \left( \frac{k}{ z}  R_0 \right)^m \, r^m \sum_{n=1}^N e^{-\iota 2\pi \frac{n}{N}  l} \cos^m(2\pi \frac{n}{N}  - \phi) \nonumber \\ \label{somma}
\end{eqnarray}

We should now try to understand which is the first relevant power of $r$, at which we can truncate the power-series expansion. It is known that, in general, $\cos^m ( 2\pi \frac{n}{N}  - \phi) = \sum_{p=0}^m C_{p,m} \cos (  2\pi \frac{n}{N} p  -  \phi p)$ where several $C_{p,m}$ may be vanishing, but $C_{m,m} \neq 0$. This means:
\begin{eqnarray}
\fl
u(r,\phi, z)  \simeq u'_0(z) e^{-\iota k \frac{r^2}{2 z}} \sum_{m=0}^\infty  \frac{ \iota^m }{m!} \left( \frac{k}{ z}  R_0 \right)^m \, r^m \sum_{n=1}^N e^{\iota 2\pi \frac{n}{N}  l}  \sum_{p=0}^m C_{p,m} \cos (  2\pi \frac{n}{N} p  -  \phi p) = \nonumber \\ \fl
= u'_0(z) e^{-\iota k \frac{r^2}{2 z}} \sum_{m=0}^\infty  \frac{ \iota^m }{m!} \left( \frac{k}{ z}  R_0 \right)^m \, r^m  \sum_{p=0}^m C_{p,m} S_{l,p} 
\end{eqnarray}
where:
\begin{eqnarray}
\fl
S_{l,p} = \sum_{n=1}^N e^{\iota 2\pi  \frac{n}{N}l} \cos (  2\pi \frac{n}{N} p -  \phi p) = 
 \sum_{n=1}^N  e^{\iota 2\pi \frac{n}{N}  l} \frac{e^{\iota \left( 2\pi \frac{n}{N} p - \phi p\right) } + e^{-\iota \left( 2\pi \frac{n}{N} p - \phi p \right)} }{2} = \nonumber \\ 
= \frac{1}{2} \left[ e^{-\iota p \phi} \sum_{n=1}^N e^{\iota 2\pi \frac{n}{N}  (l + p)} + e^{\iota p \phi} \sum_{n=1}^N e^{\iota 2\pi \frac{n}{N}  (l-p)}\right] \label{sommaTrigo} 
\end{eqnarray}
It is easy to show that $\sum_{n=1}^N e^{\iota \frac{n}{N} 2\pi (l \pm p)} = 0$ is vanishing except in the case $l=\mp p$. Therefore:
\begin{equation}
\left \lbrace
	\begin{array}{ll}
		S_{l,p} = \frac{N}{2} e^{\iota p \phi} & \mathrm{if} \,\, l = p,\\
		S_{l,p} = \frac{N}{2} e^{-\iota p \phi} & \mathrm{if} \,\, l = -p,\\
		S_{l,p} = 0  & \mathrm{else} \\
	\end{array}
\right.
\end{equation} 

In conclusion, the first relevant power of $m$ in \eref{somma}, i.e. the smallest one that is multiplied by a nonvanishing coefficient, is $m=|l|$. In fact, it is for that power of $m$ that first appear a sum of the kind  \eref{sommaTrigo} with $l=\mp p$.
Hence, by collecting all constants in $u''_0$, we can write:
\begin{equation}
u(r,\phi,z) \simeq u''_0(z)  e^{-\iota k \frac{r^2}{ 2 z}} r^{|l|}  e^{\iota  l \phi} 
\end{equation}

\end{document}